\documentstyle[12pt] {article}
\input amssym.def
\input amssym
\newfont{\Small}{cmmi7}
\newfont{\Vsmall}{cmmi5}
\newfont{\Smallcal}{cmsy7}
\newfont{\Vsmallcal}{cmsy5}
\newcommand{\Beq}{\begin{equation}}
\newcommand{\Eeq}{\end{equation}}
\newcommand{\Beqa}{\begin{eqnarray}}
\newcommand{\Eeqa}{\end{eqnarray}}
\newcommand{\Barr}{\begin{array}}

\newcommand{\Bit}{\begin{itemize}}
\newcommand{\Eit}{\end{itemize}}

\def\xz{\times}
\def\a{\alpha}\def\adt{\dot\alpha}
\def\b{\beta}\def\bdt{\dot\beta}
\def\c{\gamma}\def\C{\Gamma}
\def\d{\delta}\def\D{\Delta}
\def\e{\epsilon}
\def\F{\Phi}

\def\L{\Lambda}
\def\m{\mu}

\def\s{\sigma}\def\S{\Sigma}

\def\th{\theta}

\def\der{\partial}

\begin{document}

\title{A note on chiral fermions and heterotic strings}

\author{P.S. Howe\\ Department of Mathematics\\King's College,
London}

\maketitle

\begin{abstract}
An off-shell, manifestly $(8,0)$ worldsheet supersymmetric,
formulation of a
multiplet describing physical chiral fermions is given. The multiplet
can be
used to complete the doubly supersymmetric (twistor-like) action for
the
heterotic string.
\end{abstract}

\vfill\eject

There has been a considerable amount of interest recently in
reformulating the
actions for supersymmetric extended objects with both world-surface
and
spacetime supersymmetry. This has followed on from the initial
observation of
Sorokin et al \cite{STV89, STVZ89} that the Brink-Schwarz \cite{BS81}
superparticles in three and four dimensional spacetimes admit such a
reformulation with local worldline supersymmetry replacing the
fermionic
$\kappa$-symmmetry \cite{Sieg83} of the BS version. A number of
authors have
contributed to the development of this point of view, first to
superparticles
in higher dimensional spacetimes, and then to heterotic strings,
membranes and
higher-dimensional extended
objects
\cite{DS91,DS92,DGS92,DIS92,Ton91,Ton92,APT92,GS92,HT91,HW92,PT93a,BS9
3}. In
addition, this new approach has much in common with various
twistor-like
formulations of superparticles and  extended supersymmetric
objects\cite{BBCL87,BC88,Berk89,Berk91,Berk92a,Berk92b}. One of the
most
interesting objects to study is undoubtedly the heterotic string in
ten-dimensional spacetime. Up until now, a doubly supersymmetric
formulation
has been given for the geometrical (supergravity background) sector
of the
theory \cite{GHS92,DGHS93} and an action proposed for the fermionic
(Yang-Mills
background) sector \cite{PT93b}.
However, this latter action appears to be not completely satisfactory
as the
authors of \cite{PT93b} have themselves pointed out. It is the
purpose of this
letter to propose a replacement for this action which istself has
some
drawbacks as we shall see.

The multiplet in question is in fact known, and has appeared in the
literature
in different guises, for example, see \cite{BSN87,Berk90,GHT93}. It
is an 8+8
component multiplet which in the current context consists of 8
physical
two-dimensional chiral fermions together with 8 auxiliary boson
(scalar)
fields. It is straightforward to write down the action in components
in a flat
background, and a coupling of a similar multiplet to two-dimensional
supergravity (in components) has also been constructed \cite{BSN87}.
However,
this latter model has physical bosons and chiral fermions (still 8+8
off-shell)
and so cannot be taken over directly for the present purposes. In
addition, the
spirit of doubly supersymmetric formulations requires that the
multiplet be
written down in a manifestly supersymmetric fashion, i.e. in
superfields. In
the rest of the paper we carry this out, starting with the basic
multiplet in
flat superspace. We then briefly review (8,0) supergeometry and adapt
the
multiplet to a curved background. The action we obtain is manifestly
locally
(8,0) supersymmetric and is in addition super-Weyl invariant,
although the
latter property has to be established by hand. Finally we comment on
the
multiplet in the string context where the local worldsheet
supersymmetry
appears to force restrictions on the Yang-Mills gauge group of the
spacetime
background.

Flat (8,0) superspace is the $(2\vert 8)$-dimensional (two even and
eight odd)
supermanifold $\Bbb R^{2\vert 8}$ with standard coordinates
($x^{++},\ x^{--},\
\th^{\a +}$) where the signs indicate the Lorentz transformation
properties of
the quantities concerned. The index $\a$ indicates the
eight-dimensional spinor
representation of $SO(8)$. The supercovariant derivative $D_{\a +}$
satisfies
\Beq
\{D_{\a +},D_{\b +}\}=2i \der_{++}
\Eeq
where $\der_{++}={\der\over\der x^{++}}$ and similarly for minuses.
The
multiplet of interest is a fermionic superfield $\psi_{i -}$ which
satisfies
the constraint
\Beq
D_{\a +}\psi_{i -}=(\c_i)_{\a\adt} P_{\adt}
\label{multiplet}
\Eeq
where $P_{\adt}$ is a bosonic superfield. The index $i(=1,\dots 8)$
indicates
the vector representation of $SO(8)$ and ${\adt}$ the dotted spinor
representation (so that all three types of index, $\a,\ \adt,\ i$,
run from 1
to 8). Differentiating the above constraint and using the commutation
relations
satisfied by the $D$'s one finds
\Beq
D_{\a +} P_{\adt} = i(\c_i)_{\a\adt} \der_{++}\psi_{i -}
\Eeq
Further differentiation reveals that all the components of the
superfield
$\psi_{i -}$ are expressible in terms of the components $\psi_{i
-}\vert$,
$P_{\adt}\vert$ and spacetime ($\der_{++}$) derivatives of them, the
vertical
bars denoting evaluation of a superfield at $\th=0$. Hence the above
constraint
defines an off-shell multiplet with 8+8 components. The equation of
motion is
simply
\Beq
P_{\adt}=0
\Eeq
which immediately implies that $\der_{++}\psi_{i -}=0$. The component
action
which gives these equations is obviously the sum of the standard
kinetic term
for the chiral fermions together with an auxiliary term of the form
$P^2$.
However, in order to construct a superspace action it is necessary to
solve the
constraint (\ref{multiplet}). The solution is
\Beq
\psi_{i-}=\D_{i\adt(+7)} V_{\adt(-8)}
\label{solution}
\Eeq
where $V_{\adt(-8)}$ is the prepotential and
\Beqa
\D_{i\adt(+7)}&=&(\c_i)_{\a\adt}\big(\tilde D_{\a(+7)}+i D_\a
(\der_{++})^3\big)\nonumber \\
&\phantom{=}&-{i\over6}(\c_j)^{\a\adt} (\c_{ij})^{\b\c}\big(\tilde
D_{\a\b\c(+5)}\der_{++} + i D_{\a\b\c(+3)}(\der_{++})^2 \big)
\label{operator}
\Eeqa
The multiple derivatives are defined by
\Beqa
D_{\a_1\dots\a_n(+n)}&=&D_{[\a_1 +}\dots D_{\a_n] +} \nonumber\\
\tilde D_{\a_1\dots \a_m (+n)}&=&
{1\over n!}\e_{\a_1\dots \a_m\b_1\dots\b_n}
D_{\b_1\dots\b_n(+n)}
\Eeqa
where the square brackets denote antisymmetrisation and where, in the
second
equation, $m=8-n$. Since all $SO(8)$ indices are raised or lowered
using the
unit matrix there is no need to distinguish between upper and lower
indices.

The action which gives the desired equations of motion is
\Beq
S=\int d^2 x d^8\th V_{\adt(-8)} P_{\adt}
\Eeq
The Lagrangian has Lorentz weight $-8$ to compensate the Lorentz
weight of the
superspace measure. The action is invariant under the local gauge
transformations
\Beq
\d V_{\adt(-8)}=(\c_{ijk})_{\adt\a}D_{\a +}\L_{ijk(-9)}
\Eeq

In order to couple the above system to background (8,0) supergravity
we briefly
review the latter.
We now have a general $(2\vert8)$ real supermanifold $M$ with local
coordinates
$z^M=(x^m,\ \th^\m)$. The structure group of $M$ is taken to be
$SO(1,1)\times
SO(8)$ and a set of preferred frames denoted by $E_A=(E_a,\
E_{\a+})=(E_{++},E_{--},E_{\a+})=E_A{}^M \der_M$. The connection form
is
$\C_A{}^B$, where

\Beq
\begin{array}{lcl}
\C_{\a+}{}^{\b+}&=& B_\a{}^\b + \d_\a{}^\b C \\
\C_{++}{}^{++} &=& 2C\\
\C_{--}{}^{--}&=&-2C
\end{array}
\Eeq
are its non-vanishing components with $B$ and $C$ being the $SO(8)$
and
$SO(1,1)$ parts respectively. The torsion and curvature two-forms are
\Beq
\begin{array}{lcl}
T^A&=&d E^A + E^B \C_B{}^A \\
R_A{}^B&=& d\C_A{}^B + \C_A{}^C\C_C{}^B
\end{array}
\Eeq
where
\Beq
\begin{array}{lcl}
R_{\a+}{}^{\b+}&=& G_\a{}^\b + \d_\a{}^\b H\\
R_{++}{}^{++}&=& 2H\\
R_{--}{}^{--}&=& -2H
\end{array}
\Eeq
where $G$ and $H$ are the $SO(8)$ and $SO(1,1)$ curvatures. The
components of
the torsion tensor are constrained to satisfy
\Beq
T^{++}=i E^{\a+}\wedge E^{\a+};\qquad T^{--}=0
\Eeq
\Beq
T^{\a+}=E^{--}\wedge E^{++} \Psi_-^\a
\Eeq
where $E^A=(E^{++},E^{--},E^{\a+})=dz^M E_M{}^A$ are the basis
one-forms dual
to the preferred frames. The components of the curvature tensor are
given by
\Beq
\begin{array}{lcl}
H_{\a+\b+}&=&0\\
H_{\a+,++}&=&0\\
H_{\a+,--}&=& i\Psi_{\a-}\\
H_{++,--}&=& {1\over8}\nabla_{\a+} \Psi_{\a -}
\end{array}
\Eeq
and
\Beq
\begin{array}{lcl}
G_{\a+\b+,\c\d}&=& 0\\
G_{\a+,++,\c\d}&=& 0\\
G_{\a+,--,\c\d}&=& -\chi_{\a\c\d-}+2i\d_{\a[\c}\Psi_{\d]-} \\
G_{++,--,\c\d}&=& \nabla_{[\c+}\Psi_{\d-]}
\end{array}
\Eeq
In the above $\nabla_A$ denotes the covariant derivative with respect
to the
$SO(8)\times SO(1,1)$ structure group. The geometry is thus described
by two
fermionic superfields $\chi$ and $\Psi$, the former being totally
antisymmetric
on its $SO(8)$ indices. These superfields are themselves constrained
to satisfy
\Beq
\nabla_{\{\a+}\Psi_{\b\}-}= 0
\Eeq
and
\Beq
\nabla_{\a+}\chi_{\b\c\d}=
\nabla_{[\a+}\chi_{\b\c\d]-}-6i\d_{\a[\b}\nabla_{\c+}\Psi_{\d]-}
\label{constraint}
\Eeq
where the braces denote traceless symmetrisation.

The constraints are invariant under super-Weyl transformations, i.e.
local
rescalings of the frames; in terms of
\Beq
X_A{}^B=E_A{}^M\d E_M{}^A
\Eeq
and
\Beq
\F_{A,B}{}^C= E_A{}^M\d \C_{M,A}{}^B
\Eeq
the non-trivial super-Weyl transformations are given by
\Beq
\begin{array}{lcllcl}
X_{\a+}{}^{\b+} & =&\d_{\a}{}^{\b} S & X_{++}{}^{++} & =&
X_{--}{}^{--}=2S \\
X_{++}{}^{\a+} & = &-2i\nabla_{\a+}S & X_{--}{}^{\a+} & =&0\\
\F_{\a+} & =&\nabla_{\a+}S & \F_{\a+,\b\c} &
=&4\d_{\a[\b}\nabla_{\c]+}S
+\S_{\a\b\c+}\\
\F_{++} & =&\nabla_{++}S & \F_{--} & =&-\nabla_{--} S\\
\F_{++,\b\c} & =&-2i \nabla_{\a\b++}S & \F_{--,\b\c} & =&0
\end{array}
\Eeq
where $S$ is a scalar superfield parameter and $\S_{\a\b\c+}$ is
totally
antisymmetric. These parameters are constrained by
\Beq
\nabla_{\a+}\S_{\b\c\d+}=\nabla_{[\a+}\S_{\b\c\d]+}+
6\d_{\a[\b}\nabla_{\c\d]++}S
\label{parameters}
\Eeq
Note the similarity between this equation and equation
(\ref{constraint})
above. In fact equation (\ref{parameters}) can be solved by
\Beqa
\S_{\a\b\c+}&=& -2i\nabla_{\a\b\c(+3)} T_{--} +\s_{\a\b\c+}\\
S&=& \nabla_{++}T_{--}
\Eeqa
where $\s$ satisfies
\Beq
\nabla_{++}\s_{\a\b\c+}=0
\Eeq
as well as
\Beq
\nabla_{\a+}\s_{\b\c\d+}=\nabla_{[\a+}\s_{\b\c\d]+}
\Eeq
One can then write $\F_{\a+,\b\c}$ in the form
\Beq
\F_{\a+,\b\c}=\nabla_{\a+} L_{\b\c} +\s_{\a\b\c+}
\Eeq
where
\Beq
L_{\a\b}=-2i\nabla_{\a\b++}T_{--}
\label{Lorentz}
\Eeq
The component count for the parameters $S$ and $\S$ therefore
corresponds to an
unconstrained scalar superfield together with the superfield $\s$,
and this
exactly matches the component count of the degrees of freedom present
in the
supergravity fields $\Psi$ and $\chi$. Thus $M$ is locally
superconformally
flat. In particular, the parameter $\s$ can be used to eliminate the
$\Psi$-independent part of the solution for $\chi$ to equation
(\ref{constraint}), and we shall henceforth suppose that this has
been done.

The coupling of the chiral fermion multiplet to $(8,0)$ supergravity
is almost
trivial, at least as regards reparametrisations and $SO(1,1)\xz
SO(8)$. The
constraint (\ref{multiplet}) is simply made covariant:
\Beq
\nabla_{\a+}\psi_{i-} =(\c_i)_{\a\adt}P_{\adt}
\label{multiplet2}
\Eeq
Note that the $i$ index of $\psi$ is acted on by the local $SO(8)$
group in the
vector representation. This constraint is easily verified to be
consistent with
the supergravity constraints; indeed, the algebra of the left
covariant
derivatives ($\nabla_{\a+}, \nabla_{++}$) is unchanged from the flat
case.
Since these are the only derivatives which occur in the solution, it
immedately
follows that flat space solution, equation \ref{solution}, can be
immediately
generalised to the curved case provided that one replaces the
ordinary
derivatives in the differential operator (\ref{operator}) by
covariant
derivatives. The action is
\Beq
S=\int d^2 x d^8\th E V_{\adt(-8)} P_{\adt}
\Eeq
where $E$ is the superdeterminant of the supervielbein $E_M{}^A$. It
is
manifestly invariant under worldsheet superdiffeomorphisms and local
$SO(1,1)\xz SO(8)$ transformations. It is also invariant under chiral
multiplet
gauge transformations (with $\nabla$'s) and under super-Weyl
transformations
which act on the chiral multiplet by
\Beq
\d V_{\adt(-8)}= \big(-8S + 4T_{--} \nabla_{++} +2i
\nabla_{\a+}T_{--}\nabla_{\a+}\big)V_{\adt(-8)}
 + L_{\adt\bdt} V_{\bdt(-8)}
\label{Weyl1}
\Eeq
{}From this one can compute the super-Weyl variation of $\psi$ itself
to be
\Beq
\d\psi_{i-}=-S \psi_{i-} + L_{ij}\psi_{j-}-2i(\c_i)_{\a\adt}\nabla
_{\a+}T_{--} P_{\adt} +4T_{--}\nabla_{++}\psi_{i-}
\label{Weyl2}
\Eeq
while for $P$ one finds
\Beq
\d P_{\adt}=L_{\adt\bdt}P_{\bdt}
+2(\c_i)_{\adt\a}\nabla_{\a+}T_{--}\nabla_{++}\psi_{i-}+4T_{--}\nabla_
{++}P_{\adt}
\label{Weyl3}
\Eeq
where we have introduced
\Beq
\begin{array}{lcl}
L_{ij}&=&{1\over4}(\c_{ij})_{\a\b} L_{\a\b}\\
L_{\adt\bdt}&=&{1\over4}(\c_{ij})_{\adt\bdt}L_{ij}
\end{array}
\Eeq
with $L_{\a\b}$ given by (\ref{Lorentz}).
We note that equation (\ref{Weyl2}) implies that the constraint
(\ref{multiplet2}) is conformally covariant, while (\ref{Weyl3})
shows
explicitly that the equation of motion $P=0$ is invariant since it
implies that
$\nabla_{++}\psi=0$ as well.

In the string context one requires 32 chiral fermions, and therefore
four of
the above multiplets.
We combine them in a superfield $\psi_{i-}^r$ where the index $r$
runs from 1
to 4. This index can be associated with a Yang-Mills group on the
target
superspace $\underline{M}$, but this group appears to be restricted
to be at
most $SO(4)$. The constraint now reads
\Beq
\hat\nabla_{\a+}\psi_{i-}^r=(\c_i)_{\a\adt}P_{\adt}^r
\label{multiplet3}
\Eeq
where
\Beq
\hat\nabla_A=\nabla_A + A_A
\Eeq
with $A$ being the pull-back onto the world-sheet of the target space
Yang-Mills field,
\Beq
A_M=\der_M z^{\underline{M}}A_{\underline{M}}
\Eeq
and where worldsheet indices are distinguished from target space
indices by
underlining the latter. If we apply a second covariant derivative to
$\psi$ we
now find a constraint on the pull-back of the Yang-Mills field
strength tensor
$F$; it is
\Beq
F_{\a+\b+}={1\over8}\d_{\a\b}F_{\c+\c+}
\Eeq
Since we wish to consider arbitrary embeddings, this implies a
constraint on
the target space field strength tensor which is equivalent to the
equations of
motion. Indeed, one can simplify matters further by using a
conventional
constraint on the target space so that $F_{\a+\b+}=0$ on the world
sheet. Given
this, we find that ($\hat\nabla_{\a+},\hat\nabla_{++}$) obey the same
algebra
as the flat derivatives and so we can simply repeat the above
construction of
the action in this modified case.

The above constraint on the target space gauge group is not
surprising, since
similar restrictions occur for $(2,0)$ and $(4,0)$ world sheet
supersymmetries.
It may be that a way round this problem can be found; indeed, it is
not
possible to couple $E_8\times E_8$ gauge fields in the $NSR$ version
of the
heterotic string \cite{Getal85} in a straightforward manner.
\vskip 1cm
\noindent{\bf Acknowledgements.} I thank N. Berkovits and D. Sorokin
for
helpful comments.


\begin{thebibliography}{99}
\bibitem{STV89} D.P. Sorokin, V.I. Tkach and D.V. Volkov, Mod. Phys.
Lett. {\bf
A4} (1989) 901.
\bibitem{STVZ89} D.P. Sorokin, V.I. Tkach, D.V. Volkov and A.A.
Zheltukin,
Phys. Lett. {\bf 216 B} (1989) 302.
\bibitem{BS81} L. Brink and J.H. Schwarz, Phys. Lett. {\bf 100B}
(1981) 310.
\bibitem{Sieg83} W. Siegel, Phys. Lett. {\bf 128B} (1983) 397.
\bibitem{HT91} P. Howe and P. Townsend, Phys. Lett. {\bf 259B} (1991)
285.
\bibitem{DS91} F. Delduc and E. Sokatchev, Phys. Lett. {\bf 262B}
(1991) 444.
\bibitem{DS92} F. Delduc and E. Sokatchev, Class. Quant. Grav. {\bf
9} (1992)
361.
\bibitem{DGS92} F. Delduc, A. Galperin and E. Sokatchev, Nucl. Phys.
{\bf B368}
(1992) 143.
\bibitem{DIS92} F. Delduc, E.A. Ivanov and E. Sokatchev, Nucl. Phys.
{\bf B384}
(1992) 334.
\bibitem{APT92} S. Aoyama, P. Pasti and M. Tonin, Phys. Lett. {\bf
283B} (1992)
213.
\bibitem{Ton91} M. Tonin, Phys. Lett. {\bf 266B} (1991) 312.
\bibitem{Ton92} M. Tonin, Int. J. Mod. Phys. {\bf 7A} (1992) 6013.
\bibitem{GS92} A. Galperin and E. Sokatchev, Phys. Rev. {\bf D46}
(1992) 714.
\bibitem{PT93a} P. Pasti and M. Tonin, Padova preprint DFPD/93/TH/07
(1993).
\bibitem{PT93b} P. Pasti and M. Tonin, Padova preprint DFPD/93/TH/52
(1993).
\bibitem{GHS92} A.S. Galperin, P.S. Howe and K.S. Stelle, Nucl. Phys.
{\bf
B368} (1992) 248.
\bibitem{DGHS93}F. Delduc, A. Galperin, P.S. Howe and E. Sokatchev,
Phys. Rev.
{\bf D47} (1993) 578.
\bibitem{BBCL87} A. Bengtsson, I. Bengtsson, M. Cederwall and N.
Linden, Phys.
Rev. {\bf D36} (1987) 1766.
\bibitem{BC88} I. Bengtsson and M. Cederwall, Nucl. Phys. {\bf B302}
(1988) 81.
\bibitem{BSN87} E. Bergshoeff, E. Sezgin and H. Nishino, Phys. Lett.
{\bf
186B}(1987) 167.
\bibitem{BS93} E. Bergshoeff and E. Sezgin, Groningen-Texas AM
preprint CTP
TAMU-67/93, UG-5/93 (1993).
\bibitem{Berk89} N. Berkovits, Phys. Lett. {\bf 232B} (1989) 184.
\bibitem{Berk90} N. Berkovits, Phys. Lett. {\bf 241B} (1990) 497.
\bibitem{Berk91} N. Berkovits, Nucl. Phys. {\bf B350} (1991) 193.
\bibitem{Berk92a} N. Berkovits, Nucl. Phys. {\bf B379} (1992) 96.
\bibitem{Berk92b} N. Berkovits, Phys. Lett. {\bf 300B} (1992) 53.
\bibitem{GHT93} A. Galperin, P.Howe and P. Townsend, Nucl. Phys. {\bf
B402}
(1993) 531.
\bibitem{HW92} P.S. Howe and P.C. West, Int. J. Mod. Phys. {\bf A7}
(1992)
6639.
\bibitem{Getal85} D.J. Gross, J.A. Harvey, E. Martinec and R. Rohm,
Nucl. Phys.
{\bf B256} (1985) 256.





\end{thebibliography}
\end{document}